\documentclass[11pt,a4paper]{article}

\usepackage{amsmath,amssymb}
\usepackage{epsfig,graphicx}
\usepackage{subfigure}
\usepackage{graphicx}
\usepackage{rotating}
\usepackage{cancel}
\usepackage{bm}
\usepackage{axodraw4j}
\usepackage{color}
\usepackage{comment}
\usepackage{cite}

\newcommand{\chiY}{\chi_{_{Y}}}

\def\beq{\begin{equation}}
\def\eeq{\end{equation}}

\newcommand{\Aslash}{A\!\!\!/\,}
\newcommand{\Zslash}{Z\!\!\!/\,}
\newcommand{\Bslash}{X\!\!\!\!/\,}

\begin{document}
\numberwithin{equation}{section}
\title{{\normalsize  IPPP/12/94; DCPT/12/188;\hfill\mbox{}\hfill\mbox{}}\\
\vspace{2.5cm} \LARGE{ 
\textbf{LHC probes the hidden sector\\[1cm]}}}
\author{Joerg Jaeckel$^{1}$\footnote{jjaeckel@thphys.uni-heidelberg.de},
\
Martin Jankowiak$^{1}$\footnote{jankowiak@thphys.uni-heidelberg.de},
\
and Michael Spannowsky$^{2}$\footnote{michael.spannowsky@durham.ac.uk}
\\[2ex]
\small{\em $^1$ Institut f\"{u}r Theoretische Physik, Universit\"{a}t Heidelberg, Germany}\\[1.5ex]
\small{\em $^2$Institute for Particle Physics Phenomenology, Durham University, Durham DH1 3LE, UK}\\[1.5ex]
}
\date{}
\maketitle

\begin{abstract}
\noindent In this note we establish LHC limits on a variety of benchmark models for hidden sector physics using 2011 and 2012 data.
First, we consider a ``hidden'' U(1) gauge boson under which all Standard Model particles are uncharged at tree-level and which interacts with the visible sector either via kinetic mixing or higher dimensional operators. Second, we constrain scalar and pseudo-scalar particles interacting with the Standard Model via dimension five operators and Yukawa interactions, in particular including so-called axion-like particles.  In both cases we consider several different final states, including photons, electrons, muons and taus, establishing new constraints for a range of GeV to TeV scale masses.
Finally, we also comment on particles with electric charges smaller than $e$ that arise from hidden sector matter.
\end{abstract}

\newpage

\section{Introduction}
Many models of particle physics contain so-called hidden sectors. These contain particles whose interactions with Standard Model (SM) matter are much weaker than the typical gauge forces of the SM. The weakness of the interactions typically arises because SM particles are uncharged under the gauge symmetries of the hidden sector and, vice versa, the hidden sector particles are uncharged under the SM gauge symmetries.
This leaves three types of possible interactions:
\begin{itemize}
\item[(i)]{} Mixing of gauge neutral particles of the SM with neutral ones in the hidden sector. 
\item[(ii)]{} Renormalizable interactions of hidden scalars with the Higgs doublet.
\item[(iii)]{} Interactions via higher dimensional operators made from gauge singlets of SM and hidden matter.
\end{itemize}
For case (i) there is only a very limited number of options. Indeed unless we allow for right-handed neutrinos we have only a single completely gauge neutral particle in the SM: the photon or (alternatively) the hypercharge gauge boson. By Lorentz and gauge symmetry the only particle that can mix with the photon is another U(1) gauge boson. 
Similarly the only possible interactions of type (ii) are of the form $\phi^{\dagger}\phi H^{\dagger}H$, where $\phi$ is a hidden sector scalar field charged 
under hidden sector gauge groups. If $\phi$ is a gauge singlet then there is the additional possibility of the term $\phi H^{\dagger}H$.
Finally there are arbitrarily many possible interactions of type (iii), which are conveniently classified according to their dimensions.

In this note we will focus on simple test models that are popular benchmark scenarios in the search for light hidden sector particles.
With the goal of complementing existing low energy constraints we will use LHC data to extend the constraints to higher masses.  The rest of this paper is 
organized as follows.
In Sec.~\ref{sec:HP} we consider extra hidden sector U(1) gauge bosons, i.e.~``hidden photons,'' that mix with the photon/hypercharge,
also allowing for the presence of simple higher dimensional operators.
In Sec.~\ref{sec:ALP} we study (pseudo-)scalars coupled via higher dimensional operators to SM gauge boson bilinears as well as via derivative (or effective Yukawa) interactions to SM fermions.
For completeness in Sec.~\ref{sec:MCP} we review the first LHC limits on mini-charged particles, which arise from matter charged under hidden sector U(1) gauge bosons.

We note that our level of accuracy is
limited by a number of factors, including our inability to model signal efficiencies with full detector simulations and our
having to extract ATLAS and CMS data from plots. In addition, we do not include parton shower or other higher order effects. 
Consequently our exclusion limits should be understood with these limitations in mind.

\section{Hidden photons}\label{sec:HP}
\subsection{Kinetic Mixing}
Let us begin with our first test model: hidden photons.
Consider an extra U(1) gauge group. If all Standard Model particles are uncharged under this new gauge group then the dominant interaction with ordinary matter is via kinetic mixing~\cite{Holdom:1985ag} with the hypercharge U(1) gauge boson. This is encoded in the following Lagrangian,
\begin{eqnarray}
\label{LagKM}
\mathcal{L} \!\!&\supset&\!\! -\frac{1}{4} W^{a}_{\mu\nu}W^{a,\mu\nu}-\frac{1}{4} B_{\mu \nu} B^{\mu \nu}
- \frac{1}{4} X_{\mu \nu} X^{\mu \nu}
- \frac{\chiY}{2} B_{\mu \nu} X^{\mu \nu}
\\\nonumber
\!\!&+&\!\! \frac{m_{X}^2}{2} X_{\mu} X^{\mu}
+\frac{1}{2}\frac{m^{2}_{W}}{g^2}(-gW^{3}_{\mu}+g^{\prime}B_{\mu})^{2}+\frac{1}{2}m^{2}_{W}(W^{1}_{\mu}W^{1,\mu}+W^{2}_{\mu}W^{2,\mu})
\\\nonumber
\!\!&+&\!\! \mbox{SM matter and Higgs terms},
\end{eqnarray}
where $B_{\mu}$ and $W_{\mu}$ denote the usual electroweak gauge fields and
$X_\mu$ denotes the hidden U(1) field with gauge coupling $g_{_X}$. Importantly the term $\frac{\chiY}{2} B_{\mu \nu} X^{\mu \nu}$ introduces
mixing between $X_\mu$ and $B_\mu$. 

The naive one loop estimate for the mixing parameter is
\begin{equation}
\chiY \sim \frac{e g_{_X}}{6\pi^2}\log\left(\frac{m}{\Lambda}\right)
\end{equation}
where $m$ is the mass of a heavy particle coupled to both the new U(1) and hypercharge and $\Lambda$ is some cutoff scale.
In general models of field~\cite{Holdom:1985ag} and string 
theory~\cite{Dienes:1996zr,Lukas:1999nh,Abel:2003ue,Blumenhagen:2005ga,Abel:2006qt,Abel:2008ai,Goodsell:2009pi,Goodsell:2009xc,Goodsell:2010ie,Heckman:2010fh,Bullimore:2010aj,Cicoli:2011yh} a wide range of 
kinetic mixing parameters are predicted, stretching from  $\chi_{Y}\sim 10^{-12}$ to $\chi_{Y}\sim10^{-3}$.

The only coupling of the hidden photon field $X_{\mu}$ to the SM sector is via the kinetic mixing term.
To see its phenomenological consequences it is most convenient to perform two shifts,
\begin{equation}
\label{shift}
B_{\mu}\rightarrow B_{\mu}-\chiY X_\mu,\qquad {\rm followed\,\,\,by}\qquad  X_{\mu}\rightarrow \frac{1}{\sqrt{1-\chi^2_{Y}}}X_{\mu},
\end{equation}
which remove the kinetic mixing term.
Crucially, however, we now have direct couplings of the SM fields to $X_{\mu}$ as well as mixed mass terms between 
$X_{\mu}$ and $W^{3}_{\mu}$/$B_{\mu}$ that are proportional to $\chiY$. Since $\chi_{Y}$ is typically small in the following we will keep only the leading terms in $\chi_{Y}$.

The mass matrix for $B_{\mu},W^{3}_{\mu}$, and $X_{\mu}$ can now be diagonalized to obtain {\emph{three}} neutral gauge bosons.
One of these is massless and corresponds\footnote{After a suitable redefinition of the gauge couplings.} to the usual photon.
The other two are massive. For small mixing ($\chiY\!\ll\!1$ and {$|m^{2}_{W}/(m^{2}_{X}-m^{2}_{Z})|\ll 1$)
one is mostly $Z$-like, whereas the other is mostly hidden photon-like and corresponds to a new $Z^{\prime}$-like particle.  For convenience
we refer to the latter particle as the hidden photon $X$ in the following.
In the limit of small mixing the mass of $X$ is given by the hidden photon mass parameter $m_{X}$ appearing in
Eq.~\eqref{LagKM}. 
Performing the shift \eqref{shift} and going to the mass eigenstate basis the coupling of the hidden photon
to SM particles is given by
\begin{equation}
\label{charges}
Q_{Z^{\prime}}=\chiY g^{\prime}\left[\frac{\gamma}{\tan^{2}(\theta_{W})}T^{3}-(1+\gamma)Q_{Y}\right],\qquad {\rm where}\qquad
\gamma=\tan^{2}(\theta_{W})\frac{m^{2}_{W}}{m^{2}_{X}-m^{2}_{Z}}.
\end{equation} 

Both ATLAS~\cite{ATLAS:2012hh} and CMS~\cite{CMS:2012it} have searched for narrow $Z^{\prime}$-like
resonances in the electron and muon channels. 
The data are given as limits on the product of the production cross section with the branching ratio into leptons.
Using the charges given in Eq.~\eqref{charges} for the hidden photon we can calculate its production cross section and branching ratios and use the reported ATLAS and CMS limits to constrain the kinetic mixing parameter $\chiY$.\footnote{The CMS Collaboration has already interpreted their data in a related context (see ref.~\cite{CMS:2012it}),
while ref.~\cite{Frandsen:2012rk} discusses LHC and Tevatron bounds on kinetically mixed gauge bosons in the context of dark matter.}
To calculate the production cross section and branching ratios we use MadGraph5 v1.4.5~\cite{Alwall:2011uj} with the Hidden Abelian Higgs Model file generated with FeynRules~\cite{Christensen:2008py}. The resulting constraints are shown in Fig.~\ref{Fig:kinmix}, with the CMS results depicted as solid lines and
the ATLAS results depicted as dashed lines. The thin lines correspond to constraints from the decay into $\mu^{+}\mu^{-}$ pairs,
while the thick lines denote the combined limit from the $\mu^{+}\mu^{-}$ and $e^{+}e^{-}$ channels.

\begin{figure}[t!]
\centerline{\includegraphics[width=0.7\textwidth]{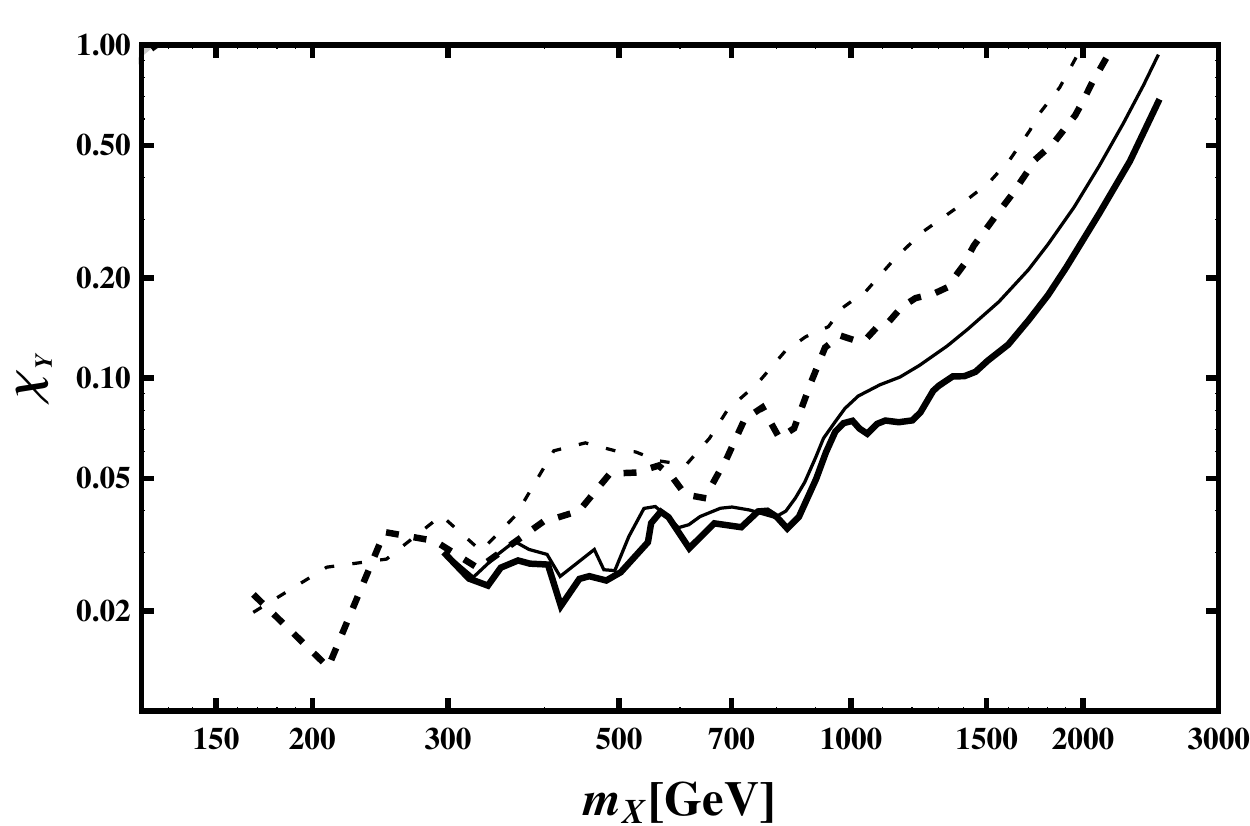}}
\caption{95\% exclusion limits on the kinetic mixing parameter $\chiY$ from the ATLAS (dashed) and CMS (solid) $Z^{\prime}$ searches. The thin lines correspond to 
the $\mu^{+}\mu^{-}$ channel only, while the thick lines result from a combination of the $\mu^{+}\mu^{-}$ and $e^{+}e^{-}$ channels.}
\label{Fig:kinmix}
\end{figure}

These new constraints extend the mass range of hidden photon tests to higher masses. This is made explicit in 
Fig.~\ref{Fig:hp_combo}, where we combine the LHC constraints (marked in orange) with a variety of other constraints.
To facilitate the comparison we have used that in the limit $m^{2}_{X}\ll m^{2}_{Z}$, which applies to the low energy bounds, the mixing of the photon with
the hidden photon, $\chi$, is related to $\chiY$ through 
\begin{equation}
\chi=\chiY\cos(\theta_{W})\qquad\qquad{\rm for}\quad m^{2}_{X}\ll m^{2}_{Z},
\end{equation}
as can be seen from Eq.~\eqref{charges}, which reduces to $Q_{Z^{\prime}}=-\chiY\cos(\theta_{W}) e [T^{3}+Q_{Y}]=-\chi e Q_{el}$
in this limit.
We can see that the LHC not only extends existing constraints to a higher mass region but that the limits are beginning
to probe quite small values of the kinetic 
mixing parameter. Nevertheless, the current limits have yet to reach the naive quantum field theory expectation of $\chiY \sim 10^{-3}$.

\begin{figure}[t!]
\centerline{\includegraphics[width=1.0\textwidth]{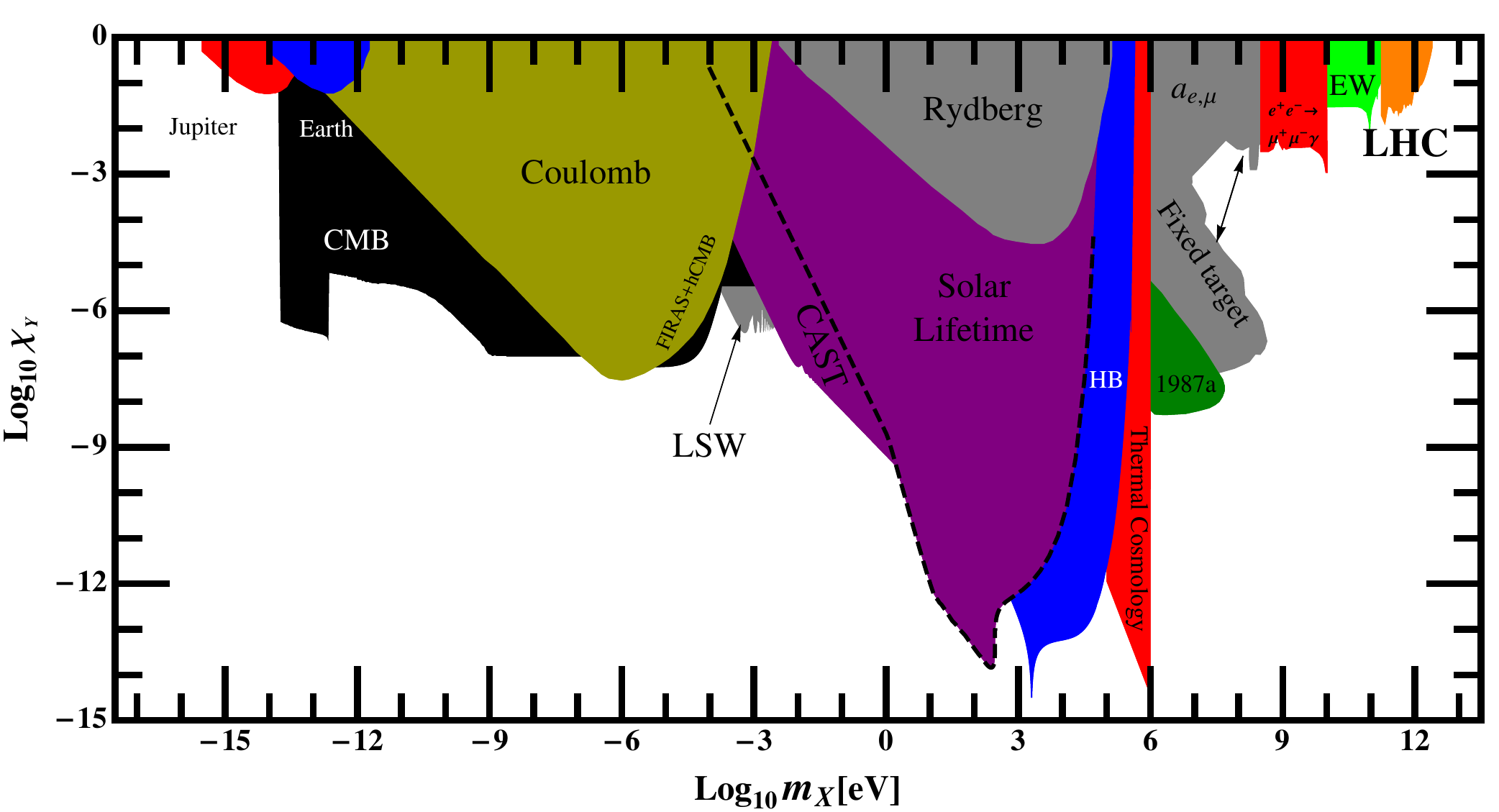}}
\caption{Combination of the new LHC limits with a range of other constraints on hidden photons (see refs.~\cite{Jaeckel:2010ni,Andreas:2012mt} for details).
The new ``LHC'' region is marked in orange and extends the existing bounds to a previously uncovered range of high masses.
Note that the limits are with respect to the hypercharge mixing parameter $\chiY$. For small hidden photon masses the kinetic mixing parameter with the ordinary photon is related to $\chiY$ through $\chi=\cos(\theta_{W})\chiY$.}
\label{Fig:hp_combo}
\end{figure}

\subsection{Dimension 6 operators}

Hidden photons can also couple to the SM via dimension 6 operators.
A full set of such operators has been collected in ref.~\cite{Dobrescu:2004wz}:
\begin{eqnarray}
\label{dim6lag}
\mathcal{L_{\rm int}} &=&
\frac{1}{M^2} \,  F_{\mu\nu}^\prime  \biggl(
C_u \overline{Q}_{\! L} \sigma^{\mu\nu}  \tilde{H} u_R
+ C_d \overline{Q}_{\! L} \sigma^{\mu\nu} H d_R
 + C_e \overline{L}_{\! L} \sigma^{\mu\nu}  H e_R + {\rm h.c.} \biggr)
\end{eqnarray}
Here $Q_L$ and $L_L$ are quark and lepton doublets, $u_R$ and  $d_R$ are up- and down-type
$SU(2)$-singlet quarks, $e_R$ are
electrically-charged $SU(2)$-singlet leptons, and $H$ is the Higgs doublet.  Sums over the three generations are left implicit.
 
Here we will not consider signals involving Higgses. Consequently we can replace the Higgs with its vev,
$\langle H\rangle =1/\sqrt{2} (0,v)^{T}$. Focusing on the simple case of universal quark and non-universal lepton couplings we have
\begin{eqnarray}
\label{dim6lagsimple}
\mathcal{L_{\rm int}} =
F_{\mu\nu}^\prime  \biggl[
\tau_{q}\,(\overline{u}_{\! L} \sigma^{\mu\nu} u_R
+ \overline{d}_{\! L} \sigma^{\mu\nu} d_R+\ldots)
 + \tau_{e}\,\, \overline{e}_{\! L} \sigma^{\mu\nu}  e_R+ \tau_{\mu}\,\, \overline{\mu}_{\! L} \sigma^{\mu\nu}  \mu_R+\ldots + {\rm h.c.} \biggr]
\end{eqnarray}
where the couplings are related to the ones in Eq.~\eqref{dim6lag} via
\begin{equation}
\tau_{q}=\frac{C_{u}}{\sqrt{2}}\frac{v}{M^2}=\frac{C_{d}}{\sqrt{2}}\frac{v}{M^2},\qquad \tau_{l_i}=\frac{C_{l_i}}{\sqrt{2}}\frac{v}{M^2}.
\end{equation}

The search strategy is essentially the same as in the previous subsection, since the hidden photon again behaves like a $Z^{\prime}$,
i.e.~like a vector-like resonance. As in the previous subsection we have calculated the cross sections with MadGraph5, using our own model file generated with FeynRules. We have then compared the resulting cross sections for the process $p p\rightarrow X\rightarrow l^{+}_{j}l^{-}_{j}$
with the exclusion limits presented in ATLAS~\cite{ATLAS:2012hh,ATLAS:2012tata} and 
CMS~\cite{CMS:2012it,Chatrchyan:2012hd}\footnote{The different structure
of the couplings in Eq.~\eqref{dim6lagsimple} as compared to the ATLAS and CMS benchmark models leads to somewhat different kinematic distributions and experimental acceptances. For the scalar case discussed in Sec.~\ref{sec:ALP} below we have checked explicitly that for the wide acceptances used
in these searches, this does not lead to dramatic differences. Nevertheless, in interpreting these limits it should be kept in mind that we are assuming
that the signal acceptances are comparable between the two cases.}. The resulting limits for the case $\tau_{q}=\tau_{l_{j}}$ are shown in Fig.~\ref{Fig:dim6}. The red, blue and green lines encode the various search channels employed: $e^{+}e^{-}$, $\mu^{+}\mu^{-}$ and $\tau^{+}\tau^{-}$. 
The gray shaded area indicates where $\Gamma_X>0.03\,m_{X}$ and one needs to take care in interpreting the limits, which
are based on searches for narrow resonances.

\begin{figure}[t!]
\centerline{\includegraphics[width=0.75\textwidth]{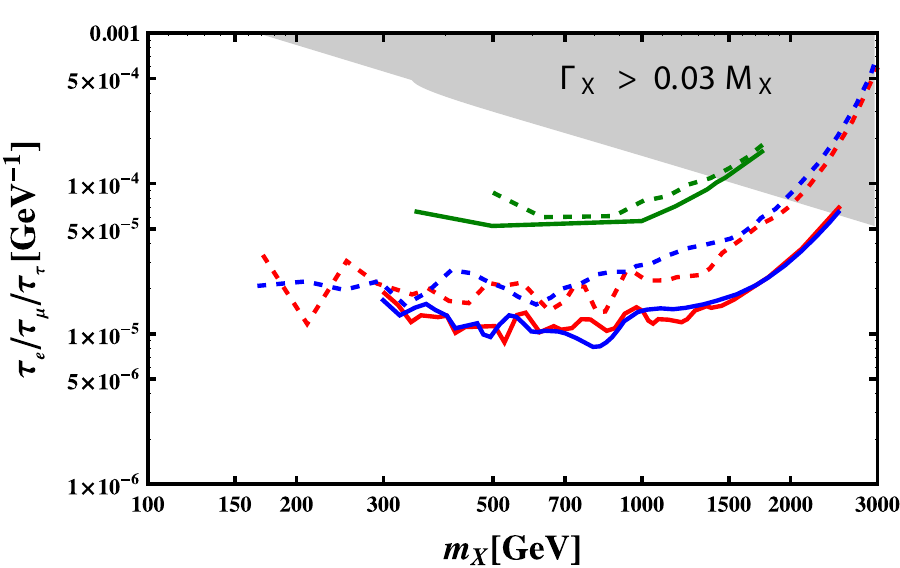}}
\caption{95\% exclusion limits on the coupling constants 
$\tau_{l_{j}}$ in the case that $\tau_{l_{j}}=\tau_{q}$ with all other lepton couplings switched off. Results
from ATLAS are shown as dashed lines, while results from CMS are shown as solid lines. Red, blue and green correspond to the $e^{+}e^{-}$, $\mu^{+}\mu^{-}$ and $\tau^{+}\tau^{-}$ channels, respectively. The gray area indicates where $\Gamma_X>0.03\,m_{X}$ and the limits need to be treated with caution. See Eq.~\eqref{dim6lagsimple} for the definition of the couplings.}
\label{Fig:dim6}
\end{figure}

Although we have shown constraints only for the specific case $\tau_{q}=\tau_{l_{j}}$ it is straightforward to repurpose the limits in Fig.~\ref{Fig:dim6}
for arbitrary ratios of the two coupling constants.
The production of the hidden photon proceeds via quark production and is therefore
controlled by $\tau_{q}^2$.
On the other hand for the branching ratio we have
$BR(X\to l^{+}_{j}l^{-}_{j})\sim\tau^{2}_{l_j}/(\sum_{i}\tau^{2}_{l_{i}}+c_{q}\tau^{2}_{q})$ with $c_{q}$ a constant that depends on $m_X$. 
Since the quark and lepton couplings have the same structure, $c_{q}$ is simply given by the number of quark species to which the decay is kinematically allowed, i.e.
\begin{equation}
c_{q}\approx 3N_{q}\approx 18\qquad\qquad {\rm for}\quad m_{X}\gg 2m_{t}.
\end{equation}
Below the top threshold $c_{q}$ is correspondingly smaller.
Thus the relevant cross section times branching ratio depends on  $\tau_{l_j}$ and $\tau_q$ as follows:
\begin{equation} 
\label{tauscaling}
\sigma_{X} \times BR(X\to l_{j}l_{j})=\sigma_{X,1}\times\frac{\tau^{2}_{q}\tau^{2}_{l,j}}{\sum_{i}\tau^{2}_{l,i}+c_{q}\tau^{2}_{q}},
\end{equation}
where in general all three lepton couplings can be switched on. Here $\sigma_{X,1}$ is the production cross section with $\tau_{q}=1$. This scaling relation can
be used to obtain limits for arbitrary ratios of $\tau_{l,j}$ and $\tau_{q}$. 
Note that since $c_{q}\approx 18\gg 1$} the case shown in Fig.~\ref{Fig:dim6} corresponds to a situation where $\sigma_{X}\!\times\! BR(X\to l_{j}l_{j})\sim (\sigma_{X,1}/c_{q})\tau^{2}_{l_{j}}$ and the limit depends on $\tau_q$ only weakly.


\section{Axion-like particles}\label{sec:ALP}
There are two possibilities for how hidden scalar and pseudo-scalar particles can interact with the SM (options (ii) and (iii) from the introduction). 
The so-called Higgs portal \cite{Patt:2006fw} is a realization of (ii).  
At very low energies and correspondingly small masses a new scalar coupled via the Higgs portal can be probed by looking for non-Newtonian ``fifth'' forces~\cite{Ahlers:2008qc}. 
At the weak scale the Higgs portal can be probed effectively in collider experiments as shown in refs.~\cite{Englert:2011us,Englert:2011yb,Englert:2011aa}. We refer the reader to these references for more details, as we will not consider this option any further in the following.

The remaining option ((iii) of the introduction) is interaction with the hidden sector via higher dimensional operators. Here there are two leading possibilities,
each of which will be considered in the following two subsections: 1) dimension 5 interactions with gauge fields; and 2) derivative or effective Yukawa couplings.

\subsection{Axion-like particles: (Pseudo-)scalars coupled to gauge boson bilinears}
Axion-like particles (ALPs) are (pseudo-)scalar particles $\phi$ of mass $m_{\phi}$ interacting with the SM through the Lagrangian
\begin{equation}
\label{twogauge} 
{\cal L}_{\phi} \supset - \frac{1}{4}\, g_{\phi BB}\, \phi\, B_{\mu \nu}\tilde{B}^{\mu \nu}
- \frac{1}{4}\, g_{\phi gg}\, \phi\, G_{\mu \nu}\tilde{G}^{\mu \nu}.
\end{equation}
Here we have written down the interaction terms
for a pseudo-scalar boson and we will continue to use this case as a benchmark in the following. 
For the scalar case one should make the replacements $\tilde{B}\rightarrow B$ and $\tilde{G}\rightarrow G$. As discussed below the LHC limits for the two cases are numerically comparable.\footnote{At low energies things are not so simple. There the differences between scalars and pseudo-scalars are enormous, as scalars contribute to fifth forces, whereas pseudo-scalars lead only to very small deviations from Newton's law. Consequently scalar interactions with the hypercharge
and color field strengths as well as first generation quarks and leptons are strongly constrained so that the pseudo-scalar case is the focus of most recent work.}

For simplicity we have included couplings only to the hypercharge U(1) and to the SU(3) field strengths, since these couplings 
are the most relevant for the signals we will study here. One could, of course, include an analogous coupling to the SU(2)
field strength. 

This form of interaction is well known from the famous axion \cite{Peccei:1977hh, Weinberg:1977ma,Wilczek:1977pj} (hence the name ALP).
In field theory it arises generally whenever a (pseudo-)scalar interacts with heavy particles
charged under the corresponding SM gauge groups.
Importantly a pseudo-scalar ALP could arise as a pseudo-Goldstone boson of some spontaneously broken symmetry and could therefore be naturally light.

In more fundamental theories, where all couplings are set by expectation values of complex scalar fields\footnote{Note that in this equation and this equation only we use a different normalization of the gauge field that is more natural for this argument.},
\begin{equation}
{\mathcal{L}}\supset -\frac{1}{4 g^{2}(\varphi)}F^{2}-\frac{\theta(\varphi)}{32\pi^2}F\tilde{F}
\end{equation}
interactions of this type naturally occur upon expanding around the vacuum expectation value
\begin{equation}
\varphi=\langle\varphi\rangle+\phi_{\rm scalar}+i\phi_{\rm pseudo-scalar}\,.
\end{equation}
For predictions from string theory see~\cite{Svrcek:2006yi,Conlon:2006tq,Arvanitaki:2009fg,Acharya:2010zx,Cicoli:2012sz}.

\subsubsection{Constraints from $\phi$-production via gluon fusion}
At the LHC the most tightly constrained signal arising from Eq.~\eqref{twogauge} is the production of $\phi$ via gluon fusion with a subsequent decay into two 
photons (a decay into two gluons, i.e.~into jets, is practically invisible above the large background).
This signal is analogous to the diphoton channel for a light Higgs, since the effective operators responsible for the production and decay of the scalar $\phi$ are the same as for the Higgs. For the pseudo-scalar case the operators include epsilon tensors, but in the highly relativistic regime applicable here the differences
between the two cases are small (see below).
Therefore for the case of a light (pseudo-)scalar $\phi$ with $m_{\phi}\in [110,150]\,{\rm GeV}$ 
we will be able to directly reinterpret the Higgs exclusion limits as constraints on $g_{\phi gg}$ and $g_{\phi BB}$.  
For the high mass region, $m_{\phi}\in [400, 2000]\,{\rm GeV}$, we will instead make direct use of ATLAS and CMS measurements of the diphoton mass spectrum that have been made in the context of extra dimension searches. 
For very low masses in the region $m_{\phi}\in [50,110]$ GeV we have made use of  ATLAS measurements of photon pair production.

\begin{figure}[t!]
\centerline{\includegraphics[width=.65\textwidth]{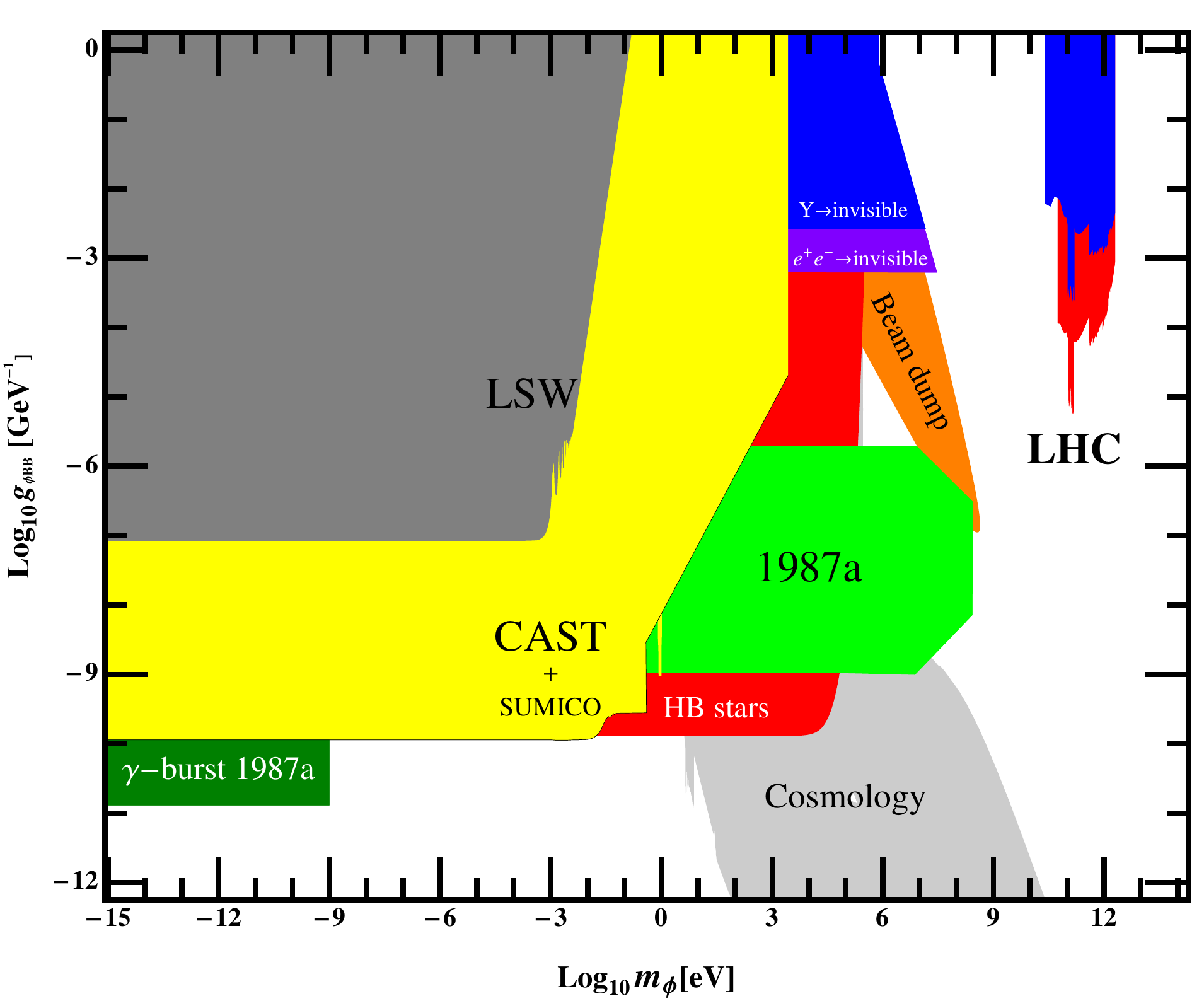}}
\caption{Summary of cosmological and astrophysical constraints
for (pseudo-)scalars coupled to two photons (compilation adapted from~\cite{Jaeckel:2010ni,Cadamuro:2011fd,Hewett:2012ns}).
The new constraints are marked in blue (pure $g_{\phi BB}$)  and red (assuming a gluon coupling with $g_{\phi gg}=g_{\phi BB}$). 
}\label{Fig:axions_combined}
\end{figure}

We have checked that the production cross sections as well as the decay widths and bulk event kinematics (at least for wide acceptances) only differ at the $\mathcal{O}(10\%)$ level between the scalar and pseudo-scalar cases. Consequently the scalar limits on $g_{\phi gg}$ and $g_{\phi BB}$ can be taken over from the pseudo-scalar case.\footnote{This is also true for most of the other constraints shown in Fig.~\ref{Fig:axions_combined}, except that in the scalar case there are some additional, stronger constraints at low masses.}

\begin{figure}[t!]
\centerline{\includegraphics[width=0.7\textwidth]{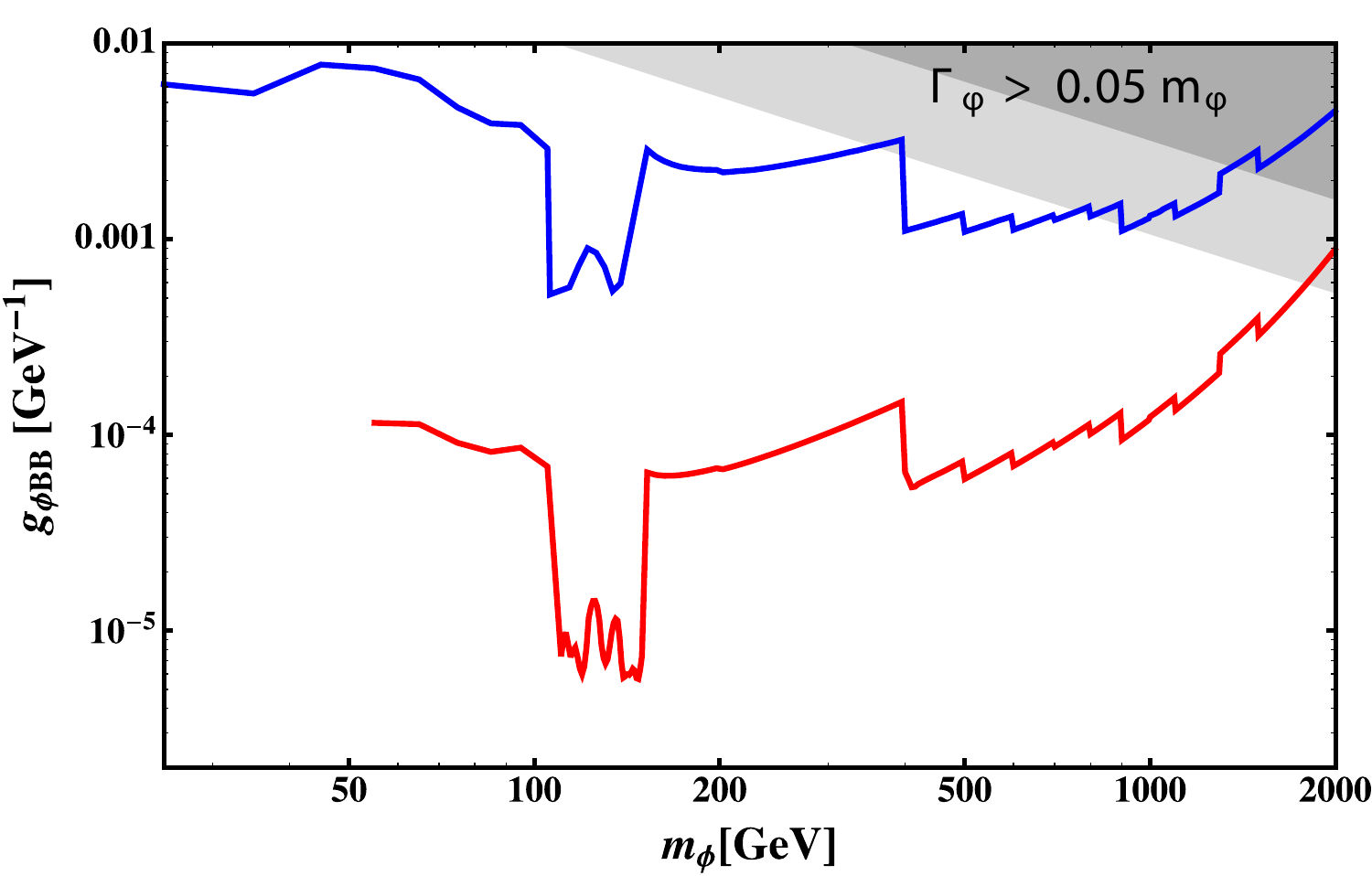}}
\caption{95\% exclusion limits on the dimension five coupling constant $g_{\phi BB}$ assuming pure photon production (blue) and
gluon production with $g_{\phi gg}=g_{\phi BB}$ (red). The limits arise from a combination of different datasets (for details see text).
The two gray regions indicate where the $\phi$ decay width $\Gamma_\phi$
exceeds $0.05\, m_\phi$ for the case of pure $g_{\phi BB}$ (dark gray) and $g_{\phi gg}=g_{\phi BB}$ (light gray). The limits need to
be interpreted with care in these regions.}
\label{Fig:higgslike}
\end{figure}

The resulting limits, which are depicted for the case of a pseudo-scalar $\phi$, are summarized in Fig.~\ref{Fig:higgslike}. The characteristic breaks where we have combined different datasets are apparent. 
In Fig.~\ref{Fig:axions_combined} we compare these LHC constraints (shown in blue and red) to a variety of other astrophysical and laboratory constraints.
We note that not only have we entered a new mass regime but that the resulting exclusion limits are relatively strong.

Let us describe how the limits were computed in greater detail.
In the mass region $m_{\phi}\in [110,150]\,{\rm GeV}$ we have used the combined results from the CMS Higgs search 
\cite{CMS-PAS-HIG-12-015}, which places a direct limit on the cross section $\sigma(H \to \gamma \gamma)$.
These limits are based on $5.1\,{\rm fb}^{-1}$ and $5.3\,{\rm fb}^{-1}$ of data taken at $E_{\rm{CM}}=7$ TeV and $E_{\rm{CM}}=8$ TeV, respectively.
It is important to note that these limits are based on NNLO cross sections.  Thus in taking over the limits directly, we are implicitly assuming that
the K-factor for the production of $\phi$ via gluon fusion is comparable to that of the Higgs.  This assumption is valid so long as the gluon fusion operator in Eq.~\eqref{twogauge} remains a good description of the physics.  For example, in the case of the Higgs this is
the case provided that the Higgs is sufficiently light, with $m_H \le 2m_t$.  To be specific, for $m_H \lesssim 150$ GeV effects due to the finite top mass are less than 5\% \cite{Anastasiou:2002yz}.  As for differences between the scalar and pseudo-scalar case, it has been shown that the K-factors only differ at the 
$\mathcal{O}(10\%)$ level for a light $\phi$ \cite{Djouadi:2005gi,Spira:1995rr,Spira:1993bb}.

To extrapolate the Higgs limits to the present case, we use the Higgs branching ratios prepared
by the LHC Higgs Cross Section Working Group~\cite{Dittmaier:2012vm}. 
The coefficient of the gluon fusion operator in the Higgs case 
is taken from ref.~\cite{Chetyrkin:1997un}.
Using these inputs we have rescaled the Higgs bounds by calculating the appropriate branching fractions and comparing them to those of the Higgs. The resulting limit, with $g_{\phi gg}$ and $g_{\phi BB}$ taken equal, is depicted in Fig.~\ref{Fig:higgslike}. 
Note that the conspicuous bump at $m_{\phi}\sim 125\, {\rm GeV}$ originates from the Higgs observation at this mass.

For the high mass region, $m_{\phi}\in [150, 2000]\,{\rm GeV}$, our limits are calculated directly from the observed number of events
in the diphoton mass spectrum as compared to the background expectation.
This is done with a Bayesian approach assuming a flat prior on the signal cross section (along the lines of ref.~\cite{Conway:2000ju}).  Specifically,
in the region $m_{\phi}\in [150, 400]\,{\rm GeV}$ we make use of $2.2\,{\rm fb}^{-1}$ of CMS data \cite{Chatrchyan:2011fq}, while for the region $m_{\phi}\in [400, 2000]\,{\rm GeV}$ we make use of $2.12 \,{\rm fb}^{-1}$ of ATLAS data \cite{ATLAS:2011ab}, both at $E_{\rm{CM}}=7$ TeV. Leading order cross sections for the diphoton signal are computed using Madgraph5 
together with a model file generated in FeynRules, with cuts and signal efficiencies implemented as in the two studies. 
The uncertainties in the signal efficiencies and integrated luminosities, which are in any case small, are not taken into account. It should be noted
that, in contrast to the low mass region, these limits are based on a LO cross section. For $\phi$ production with a K-factor greater than unity, 
these limits would be stronger by a factor of $\sqrt{K}$. This is one reason why there is a large jump in the computed exclusion limit at 
$m_{\phi} = 150\, {\rm GeV}$ (see Fig.~\ref{Fig:higgslike}).\footnote{Others include the larger integrated luminosity, lower photon $p_T$ 
requirements, and smaller $m_{\gamma\gamma}$ bins available in the low mass search.}

Finally, at very low masses $m_{\phi}\in [50,110]\,{\rm GeV}$ we have made use of the ATLAS measurements of photon pair production 
found in ref.~\cite{ATLAS:2012je}.  Again we have used a Bayesian approach to determine the maximum allowed signal cross sections in each individual mass bin,
using the predictions from $2\gamma$NNLO as the background expectation \cite{Catani:2011qz}.  The various systematic and theoretical
uncertainties are taken into account. 
The leading order signal cross sections are calculated with Madgraph5 with cuts implemented as in the ATLAS measurement,
including in particular a photon $p_T$ requirement $p_{\rm T}\geq 25\,{\rm GeV}$.
Note that although the data in ref.~\cite{ATLAS:2012je} extend below $50\,{\rm GeV}$ the cross section is vanishing at leading order for $m_\phi < 50$ GeV 
because we are not allowing for initial state radiation to give $\phi$ the transverse kick it needs in order for its decay products to (occasionally) pass the $p_T$ requirement. Although
we have not done so here, limits in this region could be established if proper care were taken to model the production process more accurately.

Let us now generalize our limits somewhat.
In Fig.~\ref{Fig:higgslike} we have let $\phi$ couple with equal strength to hypercharge and color.  It is straightforward to repurpose
this exclusion limit for arbitrary values of $g_{\phi gg}$ and $g_{\phi BB}$ using the fact that
the relevant cross section times branching ratio scales as
\begin{equation} 
\sigma_{\phi} \times BR(\phi\to \gamma\gamma)
\propto \frac{g^{2}_{\phi gg}g^{2}_{\phi BB}}{g^{2}_{\phi BB}+c_{g}g^{2}_{\phi gg}}
\label{gphibbgphiggscaling}
\end{equation}
Here the coefficient $c_{g}\approx 8$ accounts for the large 
number of gluons (at lower masses, where the Z channels are suppressed, $c_{g}$ is a bit higher).
Note that since $c_{g}$ is quite large, the case $g_{\phi gg}=g_{\phi BB}$ closely approximates the limit 
$g_{\phi gg}\gg g_{\phi BB}$ in which the exclusion limit depends only on $g_{\phi BB}$.

\subsubsection{Constraints from $\phi$-production via photon fusion}

\begin{figure}[t!]
\centerline{\includegraphics[width=.3\textwidth]{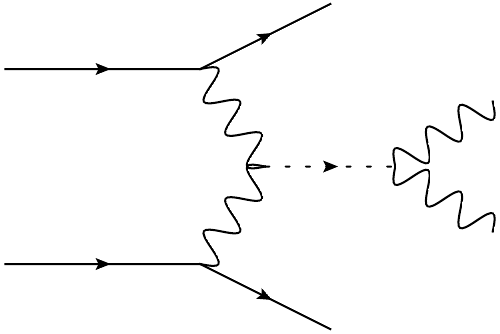}}
\caption{Feynman diagram for the production of $\phi$ for the pure $g_{\phi BB}$ case via a VBF-like topology, with $pp\to \phi+ jj\to\gamma\gamma+jj$.
}\label{Fig:ohotonprod}
\end{figure}

In the previous subsection we have seen that strong limits can be placed on $g_{\phi gg}$ and $g_{\phi BB}$ if they
are of comparable magnitude. All of the low energy constraints shown in Fig.~\ref{Fig:axions_combined}, however,
depend only on the coupling between $\phi$ and the photon. 
If we turn off $g_{\phi gg}$ it is no longer the case that $\phi$ can be produced copiously via gluon fusion.
It can, however, be produced via a VBF-like topology (see  Fig.~\ref{Fig:ohotonprod}), which allows us to establish (weaker) limits
on the pure $g_{\phi BB}$ case.  In computing these limits we proceed as before, making use of the same datasets for
$m_\phi < 100$ GeV and $m_\phi > 160$ GeV as in the gluon fusion case.\footnote{Note that because, in contrast to above, we are now
considering a four particle final state, the $p_T$ distributions of the $\phi$ decay products are now such that the photon fusion limit extends below
$m_\phi = 50$ GeV.} The resulting limits are shown in blue in Fig.~\ref{Fig:higgslike}.

In the region $m_{\phi}\in [100,160]\,{\rm GeV}$ we have used VBF data from the ATLAS Higgs search~\cite{ATLAS-CONF-2012-091,:2012gk},
establishing the maximum allowed cross sections in each mass bin with the same Bayesian approach as above.  Since this search requires two
forward jets, we get much stronger constraints in this mass region than we do for $m_\phi < 100$ GeV and $m_\phi > 160$ GeV, where the data
are inclusive.  In computing the cross sections, we do a parton level analysis and apply the same VBF cuts as in the ATLAS study.  We also assume a (conservative) signal efficiency of 50\%.
Although the resulting photon fusion bounds are weaker than the gluon fusion bounds, they have the advantage that they apply to the pure $g_{\phi BB}$ case.

\subsection{(Additional) Derivative couplings to SM fermions}
Another possible dimension five coupling between a pseudo-scalar particle $\phi$ and the SM is through derivatives:
\begin{equation}
\label{derivativecoupling}
{\mathcal L}\supset \frac{\partial^{\mu}\phi}{M}\left[ Q_{q} \left\{\bar{Q}_{L}\gamma^{\mu}Q_{L}-\bar{u}_{R}\gamma^{\mu}u_{R}-\bar{d}_{R}\gamma^{\mu}d_{R}\right\}  + Q_{l}\left\{\bar{L}_{L}\gamma^{\mu}L_{L}-\bar{e}_{R}\gamma^{\mu}e_{R}\right\}
\right]
\end{equation}
(For the scalar case we replace the minus signs with plus signs.)
This type of coupling is typical for axion-like particles arising as pseudo-Goldstone bosons or in string theory setups~\cite{Cicoli:2012sz}.

At tree level one can use the equations of motion for the fermions.
For the pseudo-scalar case the derivative coupling in Eq.~\eqref{derivativecoupling} is then equivalent to a pseudo-scalar Yukawa interaction of strength
\begin{equation}
y_{l,q} \sim \frac{Q m}{M}
\end{equation}
where $m$ is the mass of the quark or lepton in question.
For scalars the corresponding terms vanish. 
In the following we will therefore instead directly consider scalar and pseudo-scalar Yukawa 
couplings. For the scalar case we have (after electroweak symmetry breaking):
\begin{equation}
\label{derivativecoupling}
{\mathcal L}\supset \phi\left[  \left(\kappa_{u} \bar{u}u+\kappa_{d}\bar{d}d+...\right)+\left(\kappa_{e}\bar{e}e+\kappa_{\mu}\bar{\mu}\mu+...\right)\right].
\end{equation}
For the pseudo-scalar case the fermion fields come with an additional $\gamma^{5}$. 

\begin{figure}[t!]
\centerline{\includegraphics[width=0.7\textwidth]{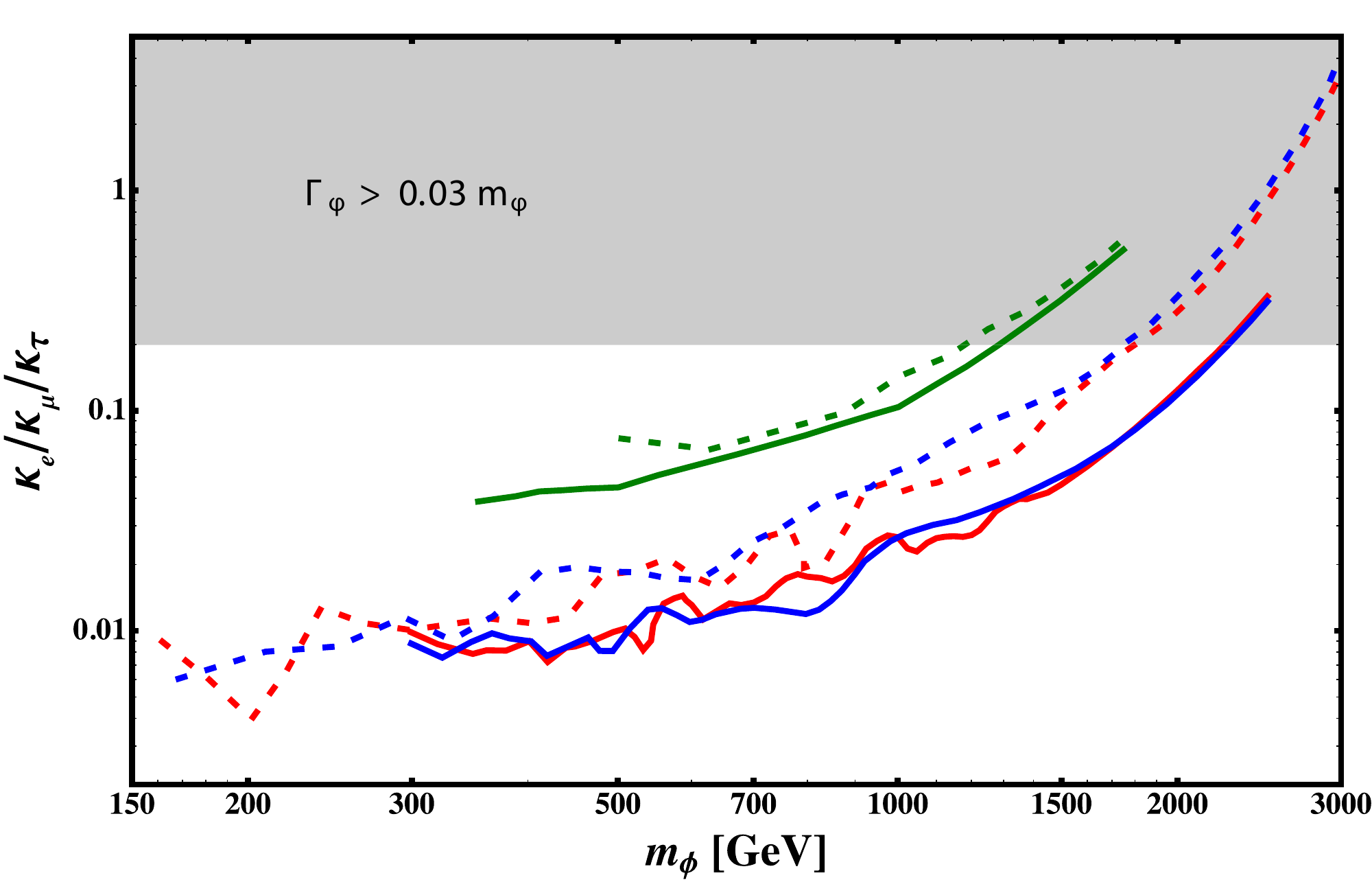}}
\caption{Constraints on the Yukawa-type coupling of $\phi$ to leptons with production via a universal coupling to quarks with $\kappa_q = \kappa_{l_i} $.
The $95\%$ exclusion limits are based on ATLAS (dashed) and CMS (solid) searches. 
The red, blue and green lines correspond to the $e^{+}e^{-}$, $\mu^{+}\mu^{-}$ and $\tau^{+}\tau^{-}$ channels, respectively.  The gray area indicates where the width of the resonance $\Gamma_\phi \geq 0.03m_{\phi}$ and the limits needs to be interpreted with care.}
\label{Fig:quarkfusion}
\end{figure}

For non-vanishing $\kappa_{l_{i}}$ (and with a reasonable branching fraction to leptons) we will again get constraints from searches for dilepton resonances. 
We use the same data and strategy as in Sec.~\ref{sec:HP}.
For a universal quark coupling $\kappa_{q}$ and with $\kappa_{q}=\kappa_{l_{j}}$ (a regime in which the limit depends only weakly on $\kappa_{q}$) the resulting limits are shown in Fig.~\ref{Fig:quarkfusion}. 
Note that since these limits are based on searches for narrow $Z^{\prime}$-like resonances (with $\Gamma \lesssim 0.03 M$), it is important to take care that the limits are not extrapolated to regions where $\phi$ becomes excessively wide.  The gray area in Fig.~\ref{Fig:quarkfusion} shows the region 
where the width $\Gamma_\phi \geq 0.03\,m_{\phi}$ and the limits need to be interpreted with care. 

\begin{figure}[t!]
\centerline{\includegraphics[width=0.7\textwidth]{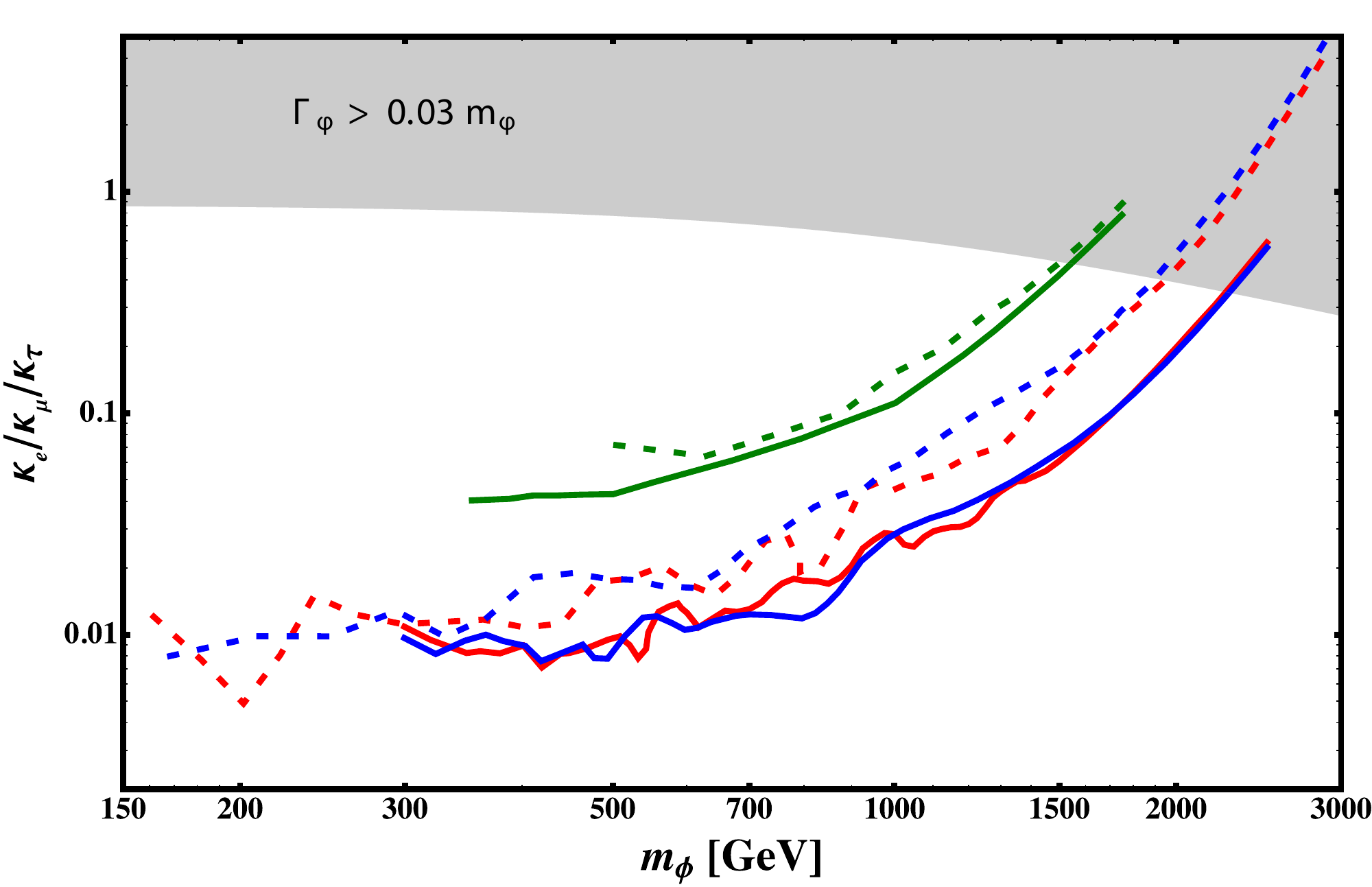}}
\caption{Constraints on the Yukawa-type coupling of $\phi$ to leptons with production via gluons and $\kappa_{l_i} = g_{\phi gg} \times \Lambda_0$ with $\Lambda_0 = 1$ TeV. The $95\%$ exclusion limits
are based on ATLAS (dashed) and CMS (solid) searches. The red, blue and green lines correspond to the $e^{+}e^{-}$, $\mu^{+}\mu^{-}$ and $\tau^{+}\tau^{-}$ channels, respectively.  The gray area indicates where the width of the resonance $\Gamma_\phi\geq 0.03\,m_{\phi}$ and the limits needs to be interpreted with care.}
\label{Fig:gluonfusion}
\end{figure}

A (pseudo-)scalar could also have additional couplings to gauge bosons as discussed in the previous section.
Let us in particular consider the couplings $g_{\phi BB}$ and $g_{\phi gg}$ from Eq.~\eqref{twogauge}.
The limits for the case of pure gluon production with decay to leptons are shown in Fig.~\ref{Fig:gluonfusion}.
 In general the relation between $\kappa_{l_{i}}$ and $g_{\phi gg}$ is highly model dependent. Here the limits are for the specific choice $\kappa_{l_i} = g_{\phi gg} \times \Lambda_0$ with $\Lambda_0$
fixed at 1 TeV. 
Again the gray region shows where $\Gamma_\phi\geq 0.03\,m_{\phi}$ and the limits need to be interpreted with care.

Following a similar strategy as in Eq.~\eqref{gphibbgphiggscaling} we can repurpose the limits in
Figs.~\ref{Fig:quarkfusion} \& \ref{Fig:gluonfusion} for general couplings $\kappa_q, \kappa_l, g_{\phi gg},$ and $g_{\phi BB}$ with the appropriate scaling relation:
\begin{eqnarray}
\label{scaling2}
\sigma_{\phi}\!\times\! BR(\phi\to l_{j}l_{j})=\sigma_{q,1}\!\left[\kappa^{2}_{q}+g_{\phi gg}^2 \chi(m_{\phi})\right] 
\!\times\!
\frac{\kappa^{2}_{l_{j}}}{\sum_i \kappa^{2}_{l_i}+\sum_i d_{q_i}\kappa^{2}_{q_i}+f_{g} m^{2}_{\phi}g_{\phi gg}^2 +f_{B} m^{2}_{\phi}g_{\phi BB}^2 }
\end{eqnarray}
Here $\sigma_{q,1}$ is the $\phi$ production cross section with universal quark couplings $\kappa_u = \kappa_d=...=\kappa_t=1$ and $g_{\phi gg}=0$.
The term in square brackets encodes the dependence on the two production mechanisms, while the ratio of coupling constants is the 
branching fraction.

The factor $\chi(m_{\phi})$ characterizes the PDF-induced difference between the quark and gluon production mechanisms.  
Numerically we find that $\chi$ increases from $\chi \approx 0.03 \rm{\;TeV}^2$ to $\chi \approx 0.11 \rm{\;TeV}^2$ as $m_{\phi}$ ranges from 150 GeV to 2000 GeV with 
\begin{equation}
\chi(m_{\phi}) \approx \left[ 0.089+0.026 \log\left(\frac{m_{\phi}}{\rm{TeV}}\right)\right ]  \rm{TeV}^2 \rm{\;\;\;for\;} 150\rm{\;GeV} \le m_{\phi} \le 2000 \rm{\;GeV\;}
\end{equation}

Let us now turn to the branching ratio. In the general case we have both dimensionless and dimensionful coupling constants.
The general form of the branching ratio can be inferred from dimensional analysis (i.e. inserting the needed factors
of $m^{2}_{\phi}$) and counting degrees of freedom.  We have checked explicitly that the given form reproduces Madgraph5 calculations.
For the light quarks $d_{q_{i}} = 3$, while for the top quark $d_t$ ranges from 0 to 3 as the decay $\phi \to t \bar{t}$ becomes relativistic. We find that as $m_{\phi}$ ranges from 200 GeV to 3000 GeV  $f_B$ ranges from $f_B \approx \frac{1}{11}$ to $f_B = \frac{1}{8}$ as the $\phi\to ZZ$ and $\phi\to Z\gamma$ channels become kinematically accessible, while $f_g$ is constant with $f_g = 1$.


\section{Minicharged particles}\label{sec:MCP}
Particles with small unquantized electric charge, often called mini- or millicharged particles (MCPs) arise in many extensions of the Standard Model.
Minicharged fermions are particularly attractive because chiral symmetry protects their masses against quantum corrections, thus making it more natural 
for them to have small masses. 
MCPs are a natural consequence of the scenario in Sec.~\ref{sec:HP} (i.e.~extra U(1) gauge groups and kinetic mixing)
in the special case that the hidden photon is massless.
In this case any matter charged under the hidden U(1) gauge group obtains a small electric charge.\footnote{Alternatively MCPs can arise in extra dimensional scenarios~\cite{Batell:2005wa} or as hidden magnetic monopoles receiving their mass from a magnetic mixing effect~\cite{Brummer:2009cs,Bruemmer:2009ky}.}
This can be easily seen as follows. 
If $X_{\mu}$ is massless a redefinition,
\begin{equation}
\label{shift2}
X^{\mu}\rightarrow X^{\mu}-\chiY B^{\mu},
\end{equation}
allows us to remove the kinetic mixing term from the Lagrangian~\eqref{LagKM} without changing any of 
the coupling terms with SM particles (apart from field/coupling renormalization).
Except for a multiplicative renormalization of the electromagnetic gauge coupling,
$e^{2}\rightarrow e^{2}/(1+\chi^2_{Y}\cos^{2}(\theta_{W})),$
the ordinary electromagnetic gauge field $A^{\mu}$ remains unaffected by this shift.
Consider now, for example, a hidden fermion $f$ charged under $X^{\mu}$.
Applying the shift~\eqref{shift} to the coupling term, we find:
\begin{equation}
g_{X}\bar{f}\Bslash\, f\rightarrow g_{X}\bar{f}\Bslash\, f-\chiY g_{X}\cos(\theta_{W})\bar{f}\Aslash\, f+\chiY g_{X}\sin(\theta_{W})\bar{f}\Zslash\, f.
\end{equation}
Since the kinetic term is now diagonal, it is clear that the particle $f$ (which was originally charged
only under U(1)$_{\rm{hidden}}$) interacts with the U(1)$_{\rm QED}$ gauge field with an apparent charge 
\begin{equation}
\epsilon e=-\chiY g_{X} \cos(\theta_{W}).
\end{equation}
From this one can also see that there is automatically a coupling to the $Z$ boson.

\begin{figure}[t!]
\centerline{\includegraphics[width=1.0\textwidth]{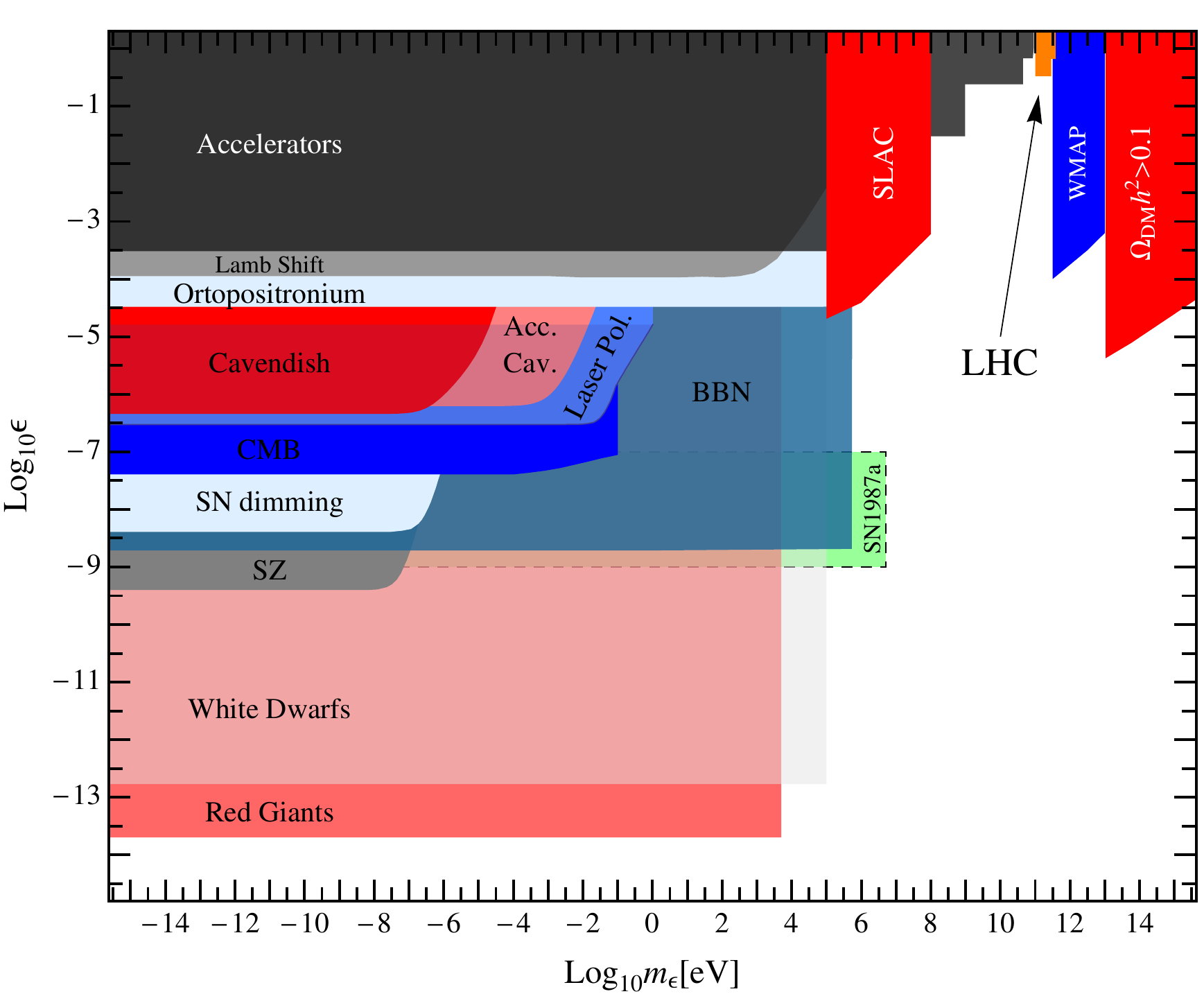}}
\caption{Combination of the new CMS limits with a range of other constraints on minicharged particles (see ref.~\cite{Jaeckel:2010ni} for details).
The new ``LHC'' region is marked in orange and extends the existing bounds to a previously uncovered mass region.}
\label{Fig:MCP}
\end{figure}

Low energy experiments as well as astrophysical and cosmological observations provide interesting constraints on MCPs. These are summarized in Fig.~\ref{Fig:MCP}.
One way to search for MCPs at the LHC would be to look for particles in the muon chamber that leave faint tracks because of their subelectronic charges.
Such an analysis has recently been performed by CMS~\cite{CMS:2012xi}. Their results are shown as the orange area in Fig.~\ref{Fig:MCP}.
One can see that the LHC fills in a gap in the region $100\,{\rm GeV}\leq m_{\epsilon} \leq 390\,{\rm GeV}$.

Alternatively we have considered the process $pp\rightarrow \mu^{+}\mu^{-}$.  The 1-loop contributions to the Z and photon propagators arising from an MCP could give rise to measurable features in the $\mu^{+}\mu^{-}$ invariant mass distribution. 
In particular such features are expected when the MCP mass crosses threshold. However, we have checked that current sensitivity 
is not sufficient to obtain new bounds.

\section{Summary and outlook}
In this note we have collected a variety of LHC results and interpreted them in terms of benchmark models of hidden sector physics with weak couplings
to the Standard Model. Among the many existing models we have focused on those that are commonly studied in the context
of low energy tests for new physics. Whereas low energy experiments provide high sensitivity at low masses, the LHC provides complementary
limits for masses in the GeV to multi-TeV range. This is particularly evident in the plots shown in Figs.~\ref{Fig:hp_combo},~\ref{Fig:MCP}
and \ref{Fig:axions_combined}, which show limits on U(1) gauge bosons coupled via kinetic mixing with hypercharge, particles with electric charges smaller than 1 and {\mbox{(pseudo-)scalars}} coupled to gauge boson bilinears, respectively. Additional results on higher dimensional couplings of extra U(1) gauge fields and (pseudo)-scalar Yukawa couplings are summarized in Figs.~\ref{Fig:dim6},~\ref{Fig:quarkfusion}~and~\ref{Fig:gluonfusion}.
The scaling expressions given in Eqs.~\eqref{tauscaling}, \eqref{gphibbgphiggscaling}, and \eqref{scaling2} allow for the various
limits to be specialized to different scenarios.

Importantly we see that in a number of different cases the LHC can probe couplings much weaker than the order unity couplings characteristic
of visible sector interactions, e.g.~kinetic mixing parameters $\chi_{\rm Y}\ll 1$ as well as Yukawa couplings $\kappa \ll 1$. Thus it is fair
to say that the LHC has begun to probe interesting regimes of hidden sector theory space
where such small couplings to Standard Model particles are expected.

Finally, there is much more data to come from the LHC. Notably, with $\sqrt{s}=14$ TeV the mass reach will be pushed higher. 
For the small couplings we are interested in a large integrated luminosity is absolutely essential so that more running time as well as a 
possible luminosity upgrade will certainly help improve the limits. Also new analyses will become available, for example photon and lepton searches with more exclusive jet requirements. These could be helpful for detecting axion-like particles as well as hidden photons. Moreover new 
searches for resonances in the top-antitop\footnote{See the ATLAS analysis~\cite{ATLAS:2012qa}.} channel will become available, allowing for top couplings to be investigated in more detail.

\section*{Note added}
Since the completion of this manuscript both ATLAS~\cite{ATLAS:2013jma} and CMS~\cite{Chatrchyan:2012oaa} have released updated searches for di-lepton resonances based on larger quantities of data.
While they do not lead to qualitative changes in the results, limits on couplings to electrons and muons using this data would be tighter by up to 50\% and extend the mass reach upwards by about 15-20\%.

\section{Acknowledgements}{\label{sec:ack}
We would like to thank Javier Redondo for interesting discussions and helpful suggestions. MS would like to thank 
the Institut f\"{u}r Theoretische Physik for hospitality.

\bibliographystyle{h-physrev5}
\bibliography{hidden.bbl}

\end{document}